# Phase Diagram of Infinite-layer Nickelate Compounds from First- and Second-principles Calculations

Yajun Zhang,[1, 2, *] Jingtong Zhang,[3, 4] Xu He,[5] Jie Wang,[4] and Philippe Ghosez[3]

[1]Key Laboratory of Mechanics on Disaster and Environment in Western China Attached to The Ministry of Education of China, Lanzhou University, Lanzhou 730000 Gansu, PR of China

[2]Department of Mechanics and Engineering Science, College of Civil Engineering and Mechanics, Lanzhou University, Lanzhou 730000 Gansu, PR of China

[3]Theoretical Materials Physics, Q-MAT, CESAM, Université de Liége, B-4000 Liége, Belgium

[4]Department of Engineering Mechanics and Key Laboratory of Soft Machines and Smart Devices of Zhejiang Province, Zhejiang University, 38 Zheda Road, Hangzhou 310027, PR of China

[5]Institute of Condensed Matter and Nanosciences (IMCN), Louvain-la-Neuve, Belgium

**Abstract**

The fundamental properties of infinite-layer rare-earth nickelates ($R$NiO$_2$) are carefully revisited and compared with those of CaCuO$_2$ and $R$NiO$_3$ perovskites. Combining first-principles and finite-temperature second-principles calculations, we highlight that bulk NdNiO$_2$ compound are far from equivalent to CaCuO$_2$, together at the structural, electronic, and magnetic levels. Structurally, it is shown to be prone to spin-phonon coupling induced oxygen square rotation motion, which might be responsible for the intriguing upturn of the resistivity. At the electronic and magnetic levels, we point out orbital-selective Mott localization with strong out-of-plane band dispersion, which should result in the isotropic upper critical fields and weakly three-dimensional magnetic interactions with in-plane local moment and out-of-plane itinerant moment. We further demonstrate that as in $R$NiO$_3$ perovskites, oxygen rotation motion and rare-earth ion controlled electronic and magnetic properties can give rise in $R$NiO$_2$ compounds to a rich phase diagram and high tunability of various appealing properties. In line with that, we reveal that key ingredients of high-$T_c$ superconductor such as orbital polarization, Fermi surface, and antiferromagnetic interactions can be deliberately controlled in NdNiO$_2$ through epitaxial strain. Exploiting strain-orbital engineering, a crossover from three- to two-dimensional magnetic transition can be established, making then NdNiO$_2$ thin film a true analog of high-T$_c$ cuprates.

## I. INTRODUCTION

Infinite-layer nickelates $R$NiO$_2$ are currently attracting intense research interest owing to the discovery of superconductivity in NdNiO$_2$ [1] and PrNiO$_2$ [2]. These intriguing observations have reinvigorated theoretical and experimental investiagtions [1-26] to rationalize the puzzling mechanism and explore effective approach to improve the superconducting temperature T$_c$.

Comparing with the $R$NiO$_3$ perovskites, $R$NiO$_2$ can be viewed as a defective NdNiO$_3$ perovskite with missing apical oxygen atoms. In $R$NiO$_3$, the motion and amplitude of antiferrodistortive (AFD) rotations, electronic state, Néel temperature, and metal-insulator transition temperature vary strongly dependent on the $R$-site ionic radius $r_R$ [27,28]. Interestingly, in the $R$NiO$_2$ family, depending on the $R$-site ion, there is also a large diversity of their properties: hole doped LaNiO$_2$ remains a metallic state, while hole doped NdNiO$_2$ and PrNiO$_2$ exhibit a superconducting phase with a higher T$_c$ in NdNiO$_2$ [1,2]; another intriguing feature of NdNiO$_2$ is the unusual upturn of its resistivity at low temperature [1], which does not appear in LaNiO$_2$ [29]. These results indicate that rare-earth ion determined property should be an essential factor to differentiate the behaviors of different compounds within the $R$NiO$_2$ family like in $R$NiO$_3$.

Magnetic interactions in high-T$_c$ superconductors are commonly believed to play essential roles for the emergence of superconductivity. Moreover, the rare-earth element is a key factor affecting the AFM coupling and superconducting temperature [30,31]. Recently, the magnetic property in NdNiO$_2$ has been verified by experiments [5]. Consequently, rationalizing the origin of the magnetism and magnetic interactions as well as the evolution of magnetic interactions in $R$NiO$_2$ series is at the heart of understanding the superconductivity mechanism and controlling T$_c$. Investigations on the exchange interactions of $R$NiO$_2$ series primarily focused on the high-symmetry $P4/mmm$ phase [21,22], while structural effect received much less attention. Identifying the fundamental role of $R$-site cation and achieving a more global description of the interplay between lattice, electron, orbital, and spin degrees of freedom in $R$NiO$_2$ compounds is therefore not only of academic interest but might also reveal of practical relevance for the optimization of their superconducting properties.

More broadly, despite significant progresses have been realized so far regarding the fundamental properties of $R$NiO$_2$ [21, 22], there are still two key questions that need to be addressed. On the one hand, the temperature dependent phase diagram is still lacking, it is not only fundamental interesting but also critical for practical control of electronic and magnetic properties by strain engineering, rotation engineering, and dimensionality engineering which play important roles in $R$NiO$_3$ perovskites. On the other hand, despite it is gradually recognized that the electronic and magnetic properties of $R$NiO$_2$ are at odds with that of cuprates, there is still lack of effective approach to make $R$NiO$_2$ analogous to cuprate superconductors to try to improve the $T_c$ in infinite-layer nickelates.

In this work, the structural, electronic, and magnetic properties of $R$NiO$_2$ compounds are investigated theoretically combining first-principles calculations at zero Kelvin and second-principles calculations at finite-temperature. Our calculations reveal the presence of strong spin-phonon coupling triggered out-of-phase rotation motion of NiO$_4$ square in NdNiO$_2$, reminiscent of the rotations of BO$_6$ octahedra in ABO$_3$ perovskites. From this, the microscopic origin of the abrupt upturn of resistivity is further rationalized. Moreover, as in $R$NiO$_3$ perovskites, rare-earth cation, oxygen rotation, and strain are demonstrated to have a profound impact on the magnetic coupling strength and electronic structure and a completely temperature-structural-magnetic phase diagram is achieved. Based on our knowledge, the spin splitting, orbital polarization and magnetic corrections are further controlled by strain engineering, yielding strained NdNiO$_2$ thin films that show marked electronic and magnetic similarities with CaCuO$_2$. So, although infinite layer nickelates and cuprates are shown to exhibit striking differences at the bulk level, our work reveal the possibility to make them similar in thin films from strain engineering, providing so a potential pathway to achieve rational control of superconductivity.

## II. STRUCTURAL PROPERTIES

**A. Spin-phonon coupling triggered rotation motion in NdNiO$_2$.**

We start our study by a careful re-investigation of the structural properties of $R$NiO$_2$ compounds ($R$ = La, Pr, Nd, Sm, Eu, Gd, Tb, Dy, Er, Tm, and Lu), relaxing first their *P*4/*mmm* phase. First-principles density functional theory (DFT) results reported in this Section were obtained using the generalized gradient approximation (GGA) [32] but are similar to those obtained using more advanced GGA + U [33] or strongly constrained and appropriately normed (SCAN) [34] as discussed in the following. The calculations have been performed with a G-type antiferromagnetic order (G-AFM) which was checked to be the GGA magnetic ground state for all compounds, consistently with previous works [16]. We will discuss further in Section III and Appendix C, Fig. 13, that the magnetic ground state should be better a C-type AFM order for GGA + U and SCAN functionals but this is not affecting the present discussions on structural property.

To probe the dynamical stability of $R$NiO$_2$ in the high-symmetry *P*4/*mmm* phase, we then calculated the phonon dispersion curves of LaNiO$_2$, NdNiO$_2$, TbNiO$_2$, and LuNiO$_2$ in both the magnetic and non-magnetic phase. Surprisingly, although LaNiO$_2$ appears as dynamically stable (see Fig. 1(a)), NdNiO$_2$, TbNiO$_2$ and LuNiO$_2$ in their magnetic phases all show sizable phonon instabilities (imaginary frequencies in Figs. 1(b)-1(d)). Moreover, decreasing the size of the *R*-site cation radii when going from LaNiO$_2$ to LuNiO$_2$, the number and amplitude of the instabilities progressively increase. The dominant instability is always an $A_4^-$ mode associated to AFD out-of-phase rotation of the NiO$_4$ squares around c axis. As the cation size decreases, additional instabilities appear that are further characterized in Appendix A, Fig. 9. The progressive destabilization of oxygen square rotation motion as the size of the *R* cation decreases is reminiscent of what is observed for AFD motions of the oxygen octahedra in $R$NiO$_3$ perovskites [35,36] (see also Appendix A, Fig. 10). Note that the AFM magnetic interactions have recently been experimentally confirmed in NdNiO$_2$, the dynamically stable of NdNiO$_2$ in the non-magnetic phase (see Fig. 1(a) and Ref. [24,25] indicates that there is strong spin-phonon coupling that triggers the appearance of rotation motion.

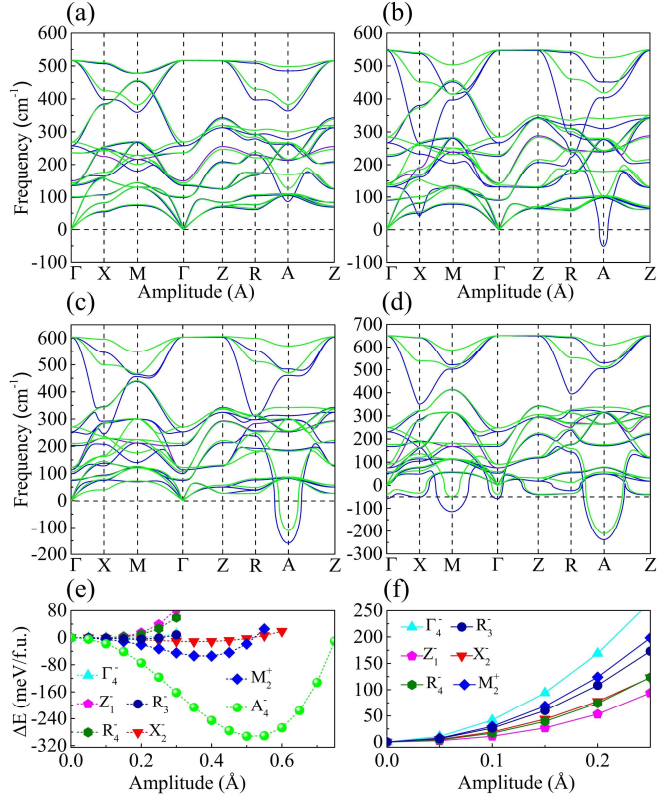

FIG. 1. Dynamical properties of $R$NiO$_2$ compounds. Phonon dispersion curves of (a) LaNiO$_2$, (b) NdNiO$_2$, (c) TbNiO$_2$, and (d) LuNiO$_2$ in their *P4/mmm* G-AFM phase (blue curve) and non-magnetic phase (green curve) as computed by GGA functional. The high-symmetry points are denoted by: $\Gamma$ = (0, 0, 0), X = (0, 0.5, 0), M = (0.5, 0.5, 0), Z = (0, 0, 0.5), R = (0, 0.5, 0.5), and A = (0.5, 0.5, 0.5). The unstable mode at high-symmetry A point is the out-of-phase rotation of the NiO$_4$ square. Potential energy surface (PES) of LuNiO$_2$ along the lines of atomic displacements corresponding to (e) the individual unstable modes at high symmetry points and (f) the same modes but with $A_4^-$ mode condensed with its natural amplitude of 0.52 Å, into the structure.

## B. The ground state structure

In order to identify the ground-state structure of $R$NiO$_2$ compounds, we fully relaxed various possible structures condensing individual and combined phonon instabilities, taking NdNiO$_2$ and LuNiO$_2$ as prototypical examples. In both cases, the identified ground state is an *I4/mcm* phase obtained from the condensation of the $A_4^-$ unstable mode of the *P4/mmm* phase.

For NdNiO$_2$, this ground state is natural since the A$_4^-$ mode is the only phonon instability. For LuNiO$_2$, the situation is more complicated since there are more unstable modes. However the A$_4^-$ instability remains dominant. As illustrated in Fig. 1(e), the double well associated the A$_4^-$ mode is significantly deeper than that related to other instabilities. Then, it is further clarified in Fig. 1(f) that, when condensing the A$_4^-$ mode with its natural amplitude, other weaker instabilities disappear (i.e. all curves switch from double- to single-well shape). This reveals an inherent competition between out-of-phase rotation and other unstable modes: the appearance of the stronger out-of-phase rotation completely suppresses the other instabilities, stabilizing the *I*4/*mcm* structure as the ground state.

Further comparison of the energy difference between *I*4/*mcm* and *P*4/*mmm* phases for all investigated *R*NiO$_2$ compounds indicates that, except for LaNiO$_2$ which remains *P*4/*mmm*, all other compounds stabilize in an *I*4/*mcm* distorted ground state. This makes *R*NiO$_2$ compounds (*R* = Pr-Lu) distinct from CaCuO$_2$ which, like LaNiO$_2$, does not exhibit any phonon instability in its *P*4/*mmm* phase. In these latter compounds, out-of-phase oxygen rotation can however be easily induced via strain engineering (see Appendix A, Fig. 11).

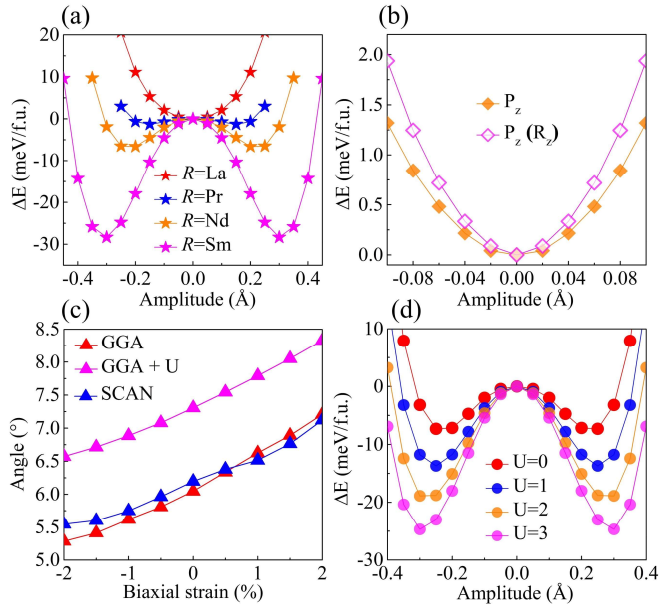

FIG. 2. Similarity between *R*NiO$_2$ and *R*NiO$_3$ from structural and chemical viewpoints. (a) PESs of out-of-phase rotation for *R*NiO$_2$ (*R* = La, Pr, Nd, and Sm), (b) PESs for the out-of-plane polarization (P$_z$) in NdNiO$_2$ with and without fixed rotation (R$_z$) of the ground state value, (c) the amplitude of rotation angle in NdNiO$_2$ as a function of epitaxial biaxial strain as obtained from GGA, GGA + U (U = 2 eV) and the strongly constrained and appropriately normed (SCAN) methods, (d) PESs of out-of-phase rotation in NdNiO$_2$ (at the experimental lattice constant) as a function of Hubbard U.

Although previously overlooked, the appearance of AFD oxygen rotation motion giving rise to an *I4/mcm* ground state is a robust result that seems to be also confirmed independently by other authors [25,26]. It appears as an intrinsic feature of the *R*NiO$_2$ infinite layer compounds that is perfectly in line with the physics of related *R*NiO$_3$ compounds, and more generally of ABO$_3$ perovskites. First, as illustrated in Figs. 2(a), the tendency to develop oxygen rotation decreases as the size of the A cation increases (Sm, Nd, Pr, La) similarly to the trend observed in the family of *R*NiO$_3$ perovskites [35]. Second, the AFD rotation motion competes together (e.g. in Fig. 1(f), the out-of-phase $A_4^-$ rotation suppresses the instability associated to in-phase $X_2^-$ rotation) as typically reported in perovskites [37] and compete also with polar distortions [38-40] (Fig. 2(b)), as also observed in YNiO$_3$ [35]. Third, oxygen rotation shows strong strain coupling like in ABO$_3$ compounds [41,42], but with an opposite trend (Fig. 2(c), AFD distortion increases under tensile biaxial strain). Finally, we see in Fig. 2 (d) that the rotation instability is strongly related to the covalency of the metal-oxygen bond which can be controlled by Hubbard U [43]: reducing the covalency through increasing the U value continuously enhances the rotation instability and the amplitude in line with perovskite oxide [43]. All these confirm that *R*NiO$_2$ infinite layer compounds show marked resemblance to that of ABX$_3$ perovskites from both structural and chemical viewpoints, providing further evidence that rotation instability should be a natural feature of ABX$_2$ compounds.

## C. Finite temperature behavior

Having established that *R*NiO$_2$ infinite layer compounds are prone to AFD distortions, the natural question that arises concerns the temperature at which those distortions appear. In ABX$_3$ perovskites, AFD distortions survive up to extremely high temperatures [35] and it might be questioned if the same is true in infinite layers.

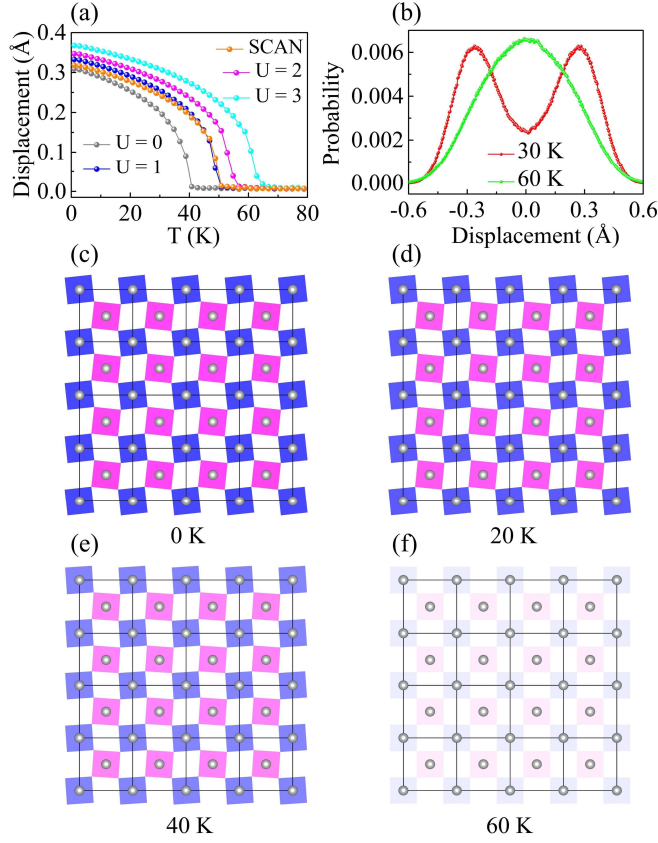

FIG. 3. Finite-temperature structural characters of $R$NiO$_2$. (a) Rotation motion induced displacement of oxygen atoms relative to their high-symmetry P4/mmm phase as a function of temperature obtained from GGA, GGA + U and SCAN methods, (b) the probability distribution of the real-space oxygen displacement in the high-symmetry P4/mmm phase below (30 K) and above (60 K) the transition temperature, and the distribution of oxygen rotation motion at (c) 0 K, (d) 20 K, (e) 40 K, and (f) 60 K. The shades of color represent the relative amplitude of rotation motion.

In order to access the finite-temperature behavior of $R$NiO$_2$ compounds, we built a second-principles model [44] with the amplitudes of individual in-plane oxygen motion along the edges of NiO$_4$ squares as degrees of freedom in the spirit of what was done by Zhong, Vanderbilt and Rabe for perovskites [45]. The model is directly fitted on first-principles data and finite-temperature properties accessed from Monte Carlo simulations. A more detailed description of the model is provided in the METHODS part and Appendix D.

Results of the Monte Carlo simulations for NdNiO$_2$ are reported in Fig. 3(a) that summarizes the temperature evolution of the average displacement associated to AFD oxygen rotation. The Figure clearly highlights a structural phase transition from the high-symmetry $P4/mmm$ phase to the low-symmetry $I4/mcm$ phase with a phase transition temperature ($T_C$) that is consistently predicted by SCAN, GGA, and GGA + U (U = 1-3 eV) at a relatively low temperature in the range from 39 to 63 K, which is not far from 70 K at which the resistance begins to increase [1].

In order to better assess the the mechanism of the phase transition and its rather displacive [46,47] or order-disorder [48] character, we have explored the probability distribution of oxygen displacements in $NdNiO_2$ with SCAN functional below (30 K) and above (60 K) the transition temperature. As displayed in Fig. 3(b), the distribution evolves from a single peak centred at zero above $T_C$ to a double peak below $T_C$, in line with a displacive character. Fig. 3(c)-(f) display the distribution of oxygen rotation motion at different temperatures and confirm this analysis. Spatially homogeneous distributions are clearly visible confirming the phases are relatively ordered at each temperature while the amplitude of the distortion is progressively reduced as the temperature increases and disappears above $T_C$. This observation of spacially homogeneous distributions is consistent with the phonon dispersion curves reported in Fig. 1(b) : the AFD insability is very localized in reciprocal space, highlighting that the appearance of the distortion requires a very large correlation volume in real space.

## III. ELECTRONIC PROPERTY
### A. Orbital-selective Mott transition

We now turn our attention to the electronic structure of $NdNiO_2$ to reveal the physical origin behind the isotropic $H_{c2}$. For that purpose, we use the SCAN functional that is expected to be more accurate and reliable than GGA (+U) for superconductors like $La_2CuO_4$ [49]. Note that all the following conclusions obtained from SCAN can be reproduced by GGA + U (see Appendix C). Consistent with previous works on $NdNiO_2$ [22], we found that C-AFM order is always the most stable for $R$ = La-Lu .

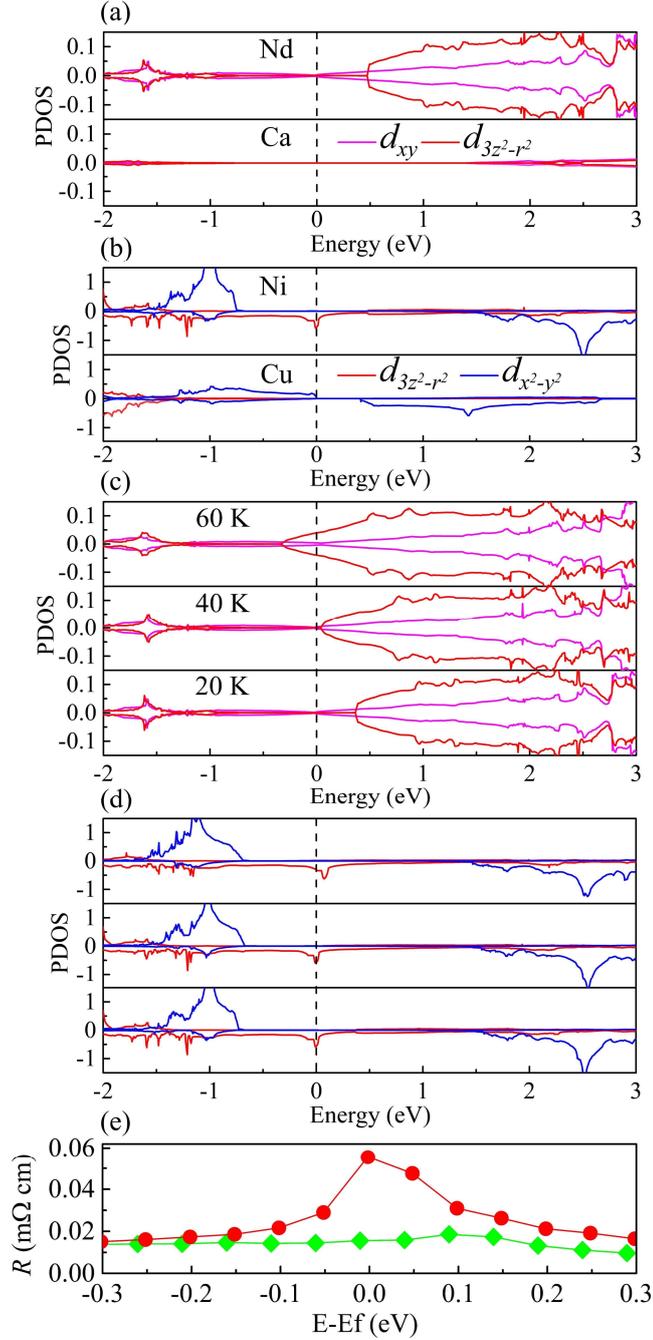

FIG. 4. Electronic structure of $CaCuO_2$ and $NdNiO_2$ and temperature dependent electronic properties of $NdNiO_2$. PDOS of (a) A-site Ca/Nd and (b) B-site Cu/Ni atoms for $CaCuO_2$ and $NdNiO_2$ in their ground state phase. The temperature determined PDOS of (c) Nd and (d) Ni atoms obtained from first-principles calculations using the structure at particular temperature extracted from our second-principles calculations. (e) the resistivity as a function of energy for the high-temperature $P4/mmm$ phase without rotation (green curve) and low-temperature $I4/mcm$ phase with rotation (red curve) is shown in Fig. The Fermi level denoted by the dashed line is set to zero energy.

The electronic states of NdNiO$_2$ near the Fermi level are dominated by the Ni $d_{x^2-y^2}$ and $d_{3z^2-r^2}$ orbitals, and Nd $d_{xy}$ and $d_{3z^2-r^2}$ orbitals consistent with prevuous work [17,22]. In Figs. 4(a) and 4(b), specific attention is paid to the projected density of states (PDOS) including the above four orbitals for CaCuO$_2$ and NdNiO$_2$. From the PDOS of CaCuO$_2$, we see clearly that the states around the Fermi level are pure Cu $d_{x^2-y^2}$ electrons with a Mott gap of 0.41 eV. However, a drastically different picture is observed in NdNiO$_2$. The correlation strengths are strongly orbital dependent: the itinerant Ni $d_{3z^2-r^2}$ orbital intersects the Fermi level while localized $d_{x^2-y^2}$ orbital lies well below the Fermi level with a Mott gap crossing the Fermi energy. This scenario confirms the orbital-selective picture in NdNiO$_2$ [17-21], and is reminiscent of the orbital-selective Mott physics that found in alkaline iron selenides [50,51] and (Ca,Sr)$_2$RuO$_4$ [52,53].

Importantly, the PDOS of NdNiO$_2$ and Fermi surface (Fig. 12) reveals a clear signature of dispersion along the k$_z$ direction, similarly to ferropnictides [54,55]. The comparable out-of-plane hopping between Ni $d_{3z^2-r^2}$ orbitals (0.46 meV) and in-plane hopping between Ni $d_{x^2-y^2}$ and O p orbitals (1.21 meV) provides another direct confirmation of the strong inter-layer coherence. It has been revealed that k$_z$ dispersion naturally facilitates the circulating currents at all field orientations and eventually results in the isotropic H$_{c2}$ as found in ferropnictides [55]. Therefore, our results support the fact that the k$_z$ dispersion in NdNiO$_2$ are likely correlated with the isotropic H$_{c2}$ as in Iron superconductors.

**B. Unusual upturn of resistivity**

Having characterized and discussed the electronic structure of the ground state structure, we explore further the electronic properties of structures at finite temperatures in order to investigate whether the predicted structural phase transition and progressive increase of rotation amplitude highlighted in Fig. 3 is the key factor that leads to the unusual upturn of resistivity of NdNiO$_2$ observed at low temperature [1].

To shed light on the actual connection between temperature and electronic structure, the PDOS of NdNiO$_2$ at different temperatures are compared in Figs. 4(c) and 4(d). Since the atomic structure appears to be spatially homogeneous at all temperatures (Figs. 3(c)-(f)), this is simply achieved by computing the DOS of homogeneous structures with the amplitude of rotation related to each temperature (Fig. 3(a)). Clearly, the PDOS in Figs. 4(c) and 4(d) reflect the fact that the electronic structure at the Fermi level is very sensitive to the amplitude of rotation, thus confirming the significant influence of temperature. In details, we found that although Nd $d_{xy}$ orbital is insensitive to temperature and rotation amplitude, the bandwidth of Nd $d_{3z^2-r^2}$ orbital decreases as the temperature decreases and the band edge shifts to an energy higher than the Fermi level, which significantly reduces the self-doping effect. The dramatic impact of temperature can also be immediately recognized by the observation that the elliptical Fermi surface of the ground state structure at the Γ point induced by the self-doping effect of Nd is strongly reduced (see Appendix B, Fig. 12) at low temperature. Additionally, the decrease of temperature is accompanied with a decrease of Ni $d_{3z^2-r^2}$ states. All these effects seem to favor a less metallic phase at low temperature, suggesting that the low-temperature resistance upturn in NdNiO$_2$ is likely to be associated with the appearance of rotation.

To clarify in practise whether rotation could be a reliable experimental fingerprint for the upturn of resistivity, a directly comparison of resistivity obtained from the Boltzmann transport equation[56] in the high-temperature *P4/mmm* phase without rotation and low-temperature *I4/mcm* phase with rotation is shown in Fig. 4(e), it is obvious that the resistivity around the Fermi level increases significantly in the low-temperature phase, confirming the important role of rotation.

## IV. MAGNETIC PROPERTIES

### A. Unified theory for weakly 3D magnetic correlations

Having elucidating the electronic structure, we further build a unified theory for the magnetic correlations in nickelate infinite-layer superconductors by a step by step analysis from the macroscopic spin-charge density to the microscopic orbital-contributed magnetic interactions.

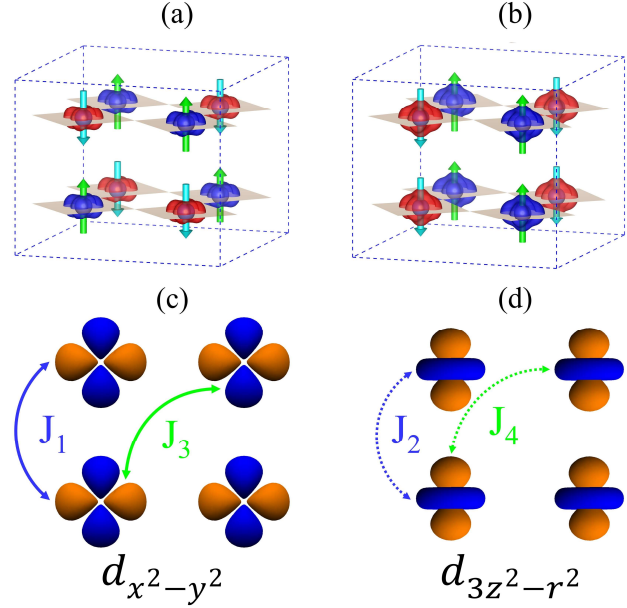

FIG. 5. Evidence of 2D and 3D magnetic interactions for $CaCuO_2$ and $NdNiO_2$, respectively. Spin-density of (a) G-AFM $P4/mmm$ $CaCuO_2$ and (b) C-AFM $I4/mcm$ $NdNiO_2$. The blue and red regions correspond to isosurfaces with spin densities of 0.02 and -0.02 $\mu B/bohr^3$. Schematic representation of exchange interactions for (c) in-plane and (d) out-of-plane magnetic coupling.

Figs. 5(a) and 5(b) illustrate the direct comparison of spin-charge density between G-AFM $CaCuO_2$ and C-AFM $NdNiO_2$. Obviously, only Cu $d_{x^2-y^2}$ electron is responsible for the magnetization of $CaCuO_2$ consistent with its 2D magnetic dimensionality [57]. In contrast, the spin-density of $NdNiO_2$ is characterized by the mixture of Ni $d_{x^2-y^2}$ and $d_{3z^2-r^2}$ electrons. The results are in full agreement with the spin splitting of the majority and minority spin orbitals shown in Fig. 4(b). Therefore, we highlight that the substantial differences between $NdNiO_2$ and $CaCuO_2$ not only lie in the electronic properties but also the magnetic order (C-AFM *vs* G-AFM) and the source of the magnetization. Interestingly, the in-plane local moment contributed by the lower-level Mott-localized electrons and the out-of-plane itinerant moment contributed by upper-level itinerant electrons again reveal striking resemblance with Iron superconductors [58-60].

Table. I. Nearest-neighbor exchange constants of NdNiO$_2$ in the unit of meV. The in-plane (J$_1$) and out-of-plane (J$_2$) nearest-neighbor exchange constants and their main orbital contributions for NdNiO$_2$.

|   | Total | $d_{x^2-y^2} - d_{x^2-y^2}$ | $d_{x^2-y^2} - d_{xy}$ | $d_{3z^2-r^2} - d_{3z^2-r^2}$ | $d_{x^2-y^2} - d_{3z^2-r^2}$ |
|---|---|---|---|---|---|
| J$_1$ | -72.5 | -66.2 | -2.4 | -0.5 | -0.4 |
| J$_2$ | 15.9 | 0 | 0 | 16.2 | 0 |

To further rationalize the origin of C-AFM order and the dimensionality of magnetic interactions in NdNiO$_2$, we develop a simple Heisenberg model and quantify the strength of exchange interactions and the orbital contributions using the TB2J code [61]. The exchange constants from different neighbors are schematically shown in Fig. 5(c) and 5(d) while the values of nearest-neighbor exchange constants and orbital contributions are listed in Table I. We see clearly that the nearest-neighbor in-plane magnetic coupling (J$_1$) mainly originates from the contribution of $d_{x^2-y^2}$ orbital, giving rise to an exchange constant of -72.5 meV which agrees well with the value of -63 meV reported in experiments [5]. Moreover, it is apparent that the presence of out-of-plane ferromagnetic coupling (J$_2$) emerges from the interactions between itinerant Ni $d_{3z^2-r^2}$ electron owning to the notable hopping between the out-of-plane nearest-neighbor Ni atoms, which facilitates the formation of weakly 3D magnetic interactions. In contrast, the $d_{3z^2-r^2}$ orbital of CaCuO$_2$ are fully occupied and not spin-polarized so that it has no contribution to the out-of-plane exchange constant, resulting in the quasi-2D AFM character. Despite SCAN normally overestimates the magnetization and magnetic energies of metals [62], our results from GGA + U also suggest that the out-of-plane magnetization and magnetic coupling are non-negligible. Consequently, the itinerant and localized electrons together conspire to account for the 3D magnetic state which is strongly analogous to Iron superconductors [58-60].

It should be noted that the SCAN and GGA + U results reveal that the in-plane neatest-neighbor exchange constant is around 10-20 times that of out-of-plane neatest-neighbor exchange constant. Thus, the out-of-plane exchange interaction is rather weak compared with in-plane magnetic interactions and maybe difficult to measure in experiments. Moreover, the different magnetization of Ni in NdNiO$_2$ (1 μB) [63] and CaCuO$_2$ (0.51 μB) [64] suggests that the electronic and magnetic properties in two compounds are different. Also, the unexpected isotropy of the upper critical field H$_{c2}$ revealed in NdNiO$_2$ also suggests that the electronic structure in NdNiO$_2$ is strongly different from CaCuO$_2$ and the k$_z$ dispersion of $d_{3z^2-r^2}$ should play critical role in the electronic and magnetic properties as in ferropnictides. Indeed, $d_{3z^2-r^2}$ is responsible for the out-of-plane magnetic coupling not only in *R*NiO$_2$, but also in *R*NiO$_{2.5}$ like LaNiO$_{2.5}$ and *R*NiO$_3$. Therefore, the predicted weakly 3D magnetic interactions in NdNiO$_2$ and its electronic origin are deserved for the experimentalist to reexamine the weakly out-out-plane magnetic coupling.

**B. Rare-earth ion dependent electron hopping and magnetic coupling**

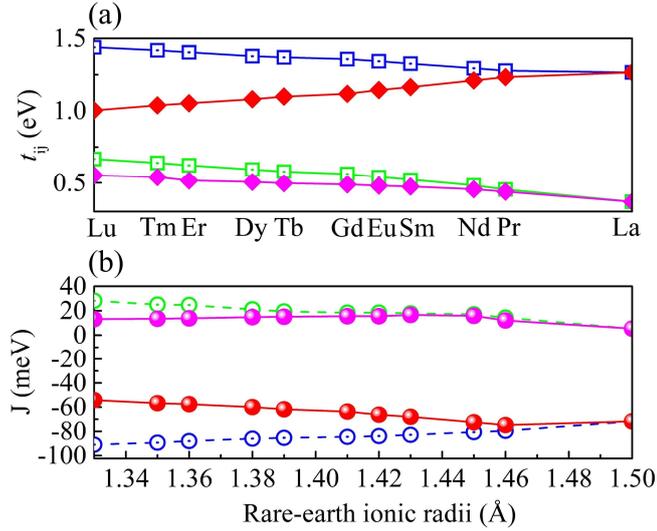

FIG. 6. *R*-site cation determined electronic and magnetic properties of *R*NiO$_2$. (a) In-plane nearest-neighbor hopping between Ni $d_{x^2-y^2}$ and O *p* orbital for C-AFM *P*4/*mmm* (blue curve) and *I*4/*mcm* (red curve) *R*NiO$_2$. The corresponding out-of-plane nearest-neighbor hopping between Ni $d_{3z^2-r^2}$ orbitals are described by blue and magenta curves. The hopping amplitude is obtained from the maximally localized Wannier function. (b) In-plane nearest-neighbor exchange constants for *P*4/*mmm* (blue curve) and *I*4/*mcm* (red curve) *R*NiO$_2$. The corresponding out-of-plane nearest-neighbor exchange constants are described by blue and magenta curves.

Typically, the orbital hybridization and magnetic coupling in ABO$_3$ perovskites strongly depend on the rotation angle: when the radii of A-site cation decreases, the lattice constant decreases while the rotation angle gradually increases. The effects of the reduction of the lattice constant and amplification of rotations are usually opposite and the actual influence of A-site cation on the electronic and magnetic properties is consequently determined by the competing effect of lattice constant and rotation. In $R$NiO$_2$, the decrease of the $R$-site cation radii also gives rise to a decrease in the lattice constant and an increase of rotation in a similar way as ABO$_3$ perovskite. It can thus be expected that the influence of lattice constant evolution is not enough to account for the global effect of $R$-site cation: the combined effect of both the lattice constant and rotation changes has to be taken into account.

In order to understand the actual role of $R$-site cation in $R$NiO$_2$ and disentangle the relative importance of lattice constant and rotation in controlling the electronic structure, the nearest-neighbor hopping parameters of $P4/mmm$ and $I4/mcm$ $R$NiO$_2$ are compared in Fig. 6(a). In the high-symmetry $P4/mmm$ phase, there is a progressive increase of both in-plane hopping between Ni $d_{x^2-y^2}$ orbital and O $p$ orbitals and out-of-plane hopping between two neighboring Ni $d_{3z^2-r^2}$ orbitals when $R$-site cation becomes smaller, owing to the decreased lattice constant and bond length. In the distorted $I4/mcm$ phases, as $r_R$ decreases, despite the out-of-plane hopping experiences a similar increase due to the reduction of Ni-Ni distance, the response of the in-plane hopping between Ni $d_{x^2-y^2}$ and O $p$ orbitals is completely opposite to the $P4/mmm$ phase, showing an intriguingly decrease. Consequently, both the lattice constant and rotation substantially affect the electronic properties of $R$NiO$_2$. In a way remarkably similar to most ABO$_3$ perovskites, the global evolution of orbital hybridization strength between B-site transition metal and oxygen is primarily associated to the rotation effect.

The variation of dominant exchange constants across the $R$NiO$_2$ series is shown in Fig. 6 (b), it is obvious that both the in-plane and out-of-plane nearest-neighbor magnetic couplings are progressively increased when decreasing $r_R$ in the $P4/mmm$ phase similarly with recent work [23]. In contrast, in-plane AFM coupling in the $I4/mcm$ phase is in contrast Ref. [23] as it gradually decreases as $r_R$ decreases. Therefore, the decreased in-plane AFM coupling for smaller $r_R$ in the ground state *I4/mcm* phase should be attributed to the increase of rotation whose effect is partly compensated by the lattice constant effect. The $r_R$ dependent AFM coupling is completely in line with rare-earth perovskites like *R*CrO$_3$ [65], *R*FeO$_3$ [66] and *R*NiO$_3$ [67], hence, the same superexchange mechanism and superexchange coupling J$\sim cos^2\theta$ developed for perovskites should also be applicable to *R*NiO$_2$. Thus, to make an accurate description of the variation of electronic and magnetic properties across the *R*NiO$_2$ series, it is crucial to consider the rotation effect like in *R*NiO$_3$.

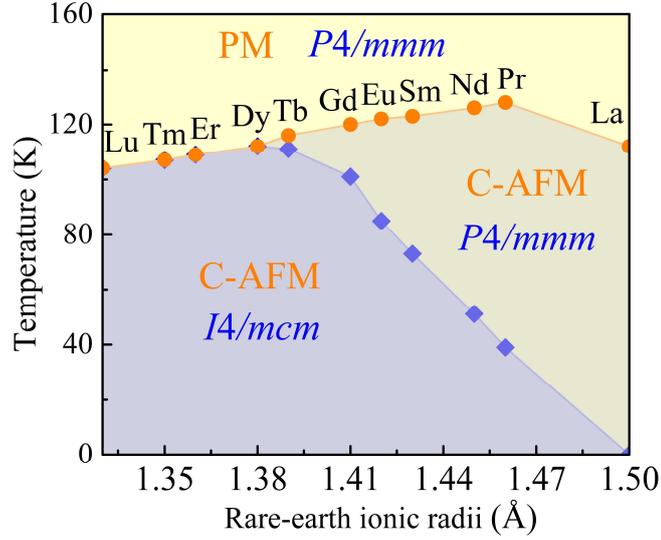

FIG. 7. Phase diagram of infinite layer nickelates in terms of rare-earth ionic radii and the temperature. The circles refer to the AFM-FM phase transition temperature and the diamonds represent the transition temperature from the I4/mcm phase to the P4/mmm phase.

## V. PHASE DIAGRAM OF RNiO$_2$ COMPOUNDS

Combining the previous Heisenberg spin Hamiltonian with the second-principles model, the temperature-dependent structural and magnetic phase transitions of the whole family of *R*NiO$_2$ compounds are determined from Monte Carlo simulations. The results are summarized in the global phase diagram reported in Fig. 7. Except for high-symmetry LaNiO$_2$, we see that all the other infinite-layer nickelate compounds undergo a structural phase transition from the low-symmetry *I4/mcm* phase to the high-symmetry *P4/mmm* phase. Moreover, as rare-earth ionic radii $r_R$ decreases, the transition temperature T$_C$ continuously increases from PrNiO$_2$ to TbNiO$_2$ due to the more distorted structure, as similarly observed in $R$NiO$_3$ perovskites. Owning to the decreased in-plane nearest-neighbor exchange constant from PrNiO$_2$ to LuNiO$_2$, it is found that the AFM-PM transition temperature T$_N$ decreases steadily as $r_R$ decreases (except the high-symmetry LaNiO$_2$) which is also strongly analogous to the behavior observed in $R$NiO$_3$ ($R$=Sm-Lu) perovskites [68]. As the predicted structural phase transition temperature for $R$=Dy-Lu in the C-AFM (PM) phase is higher (lower) than the magnetic phase transition temperature, we could deduce that the magnetic and structural phase transitions occur simultaneously for DyNiO$_2$ to LuNiO$_2$. Consequently, the presence of rotation motion not only affects the electronic property and the overall magnitude of the magnetic transition temperature, but gives rise to a much complex phase diagram with three distinct phases: (i) the C-type AFM low-symmetry *I*4*/mcm* phase, (ii) the C-type AFM high-symmetry *P*4*/mmm* phase and (iii) the PM high-symmetry *P*4*/mmm* phase. The complex phase diagram of infinite-layer nickelates show strong similarities with the phase diagram of perovskite nickelates and the dedicated interplay among the spin, electron, rotation and temperature opens up new perspectives for the control of magnetic and superconducting properties by exploiting the different strategies previously used in perovskite nickelates.

## VI. ELECTRONIC AND MAGNETIC TRANSITIONS

From the above analysis, we notice that with its $d_{3z^2-r^2}$ orbital and 3D magnetic order closely similar to iron superconductors, NdNiO$_2$ is far from being an analog of CaCuO$_2$ which shows pure single $d_{x^2-y^2}$ band and 2D AFM order. These distinct electronic and magnetic properties may be the microscopic origin of much lower T$_c$ in NdNiO$_2$. Nevertheless, if the cuprate-like characters could be engineered in NdNiO$_2$, it may become more favorable for superconductivity. As emphasized before, the key difference between NdNiO$_2$ and CaCuO$_2$ arises from the itinerant $d_{3z^2-r^2}$ electrons, which are spin-polarized and have a higher energy level than the $d_{x^2-y^2}$ electrons. Therefore, it seems plausible that pushing down the band edge of $d_{3z^2-r^2}$ orbital below the Fermi level would make NdNiO2 more analogous CaCuO2. Doing so, the $d_{3z^2-r^2}$ orbital would be nearly fully occupied similarly to CaCuO2, the 3D magnetic interactions would correspondingly be reduced to 2D AFM interactions and the states around the Fermi lever would be mainly dominated by the single Ni $d_{x^2-y^2}$ band.

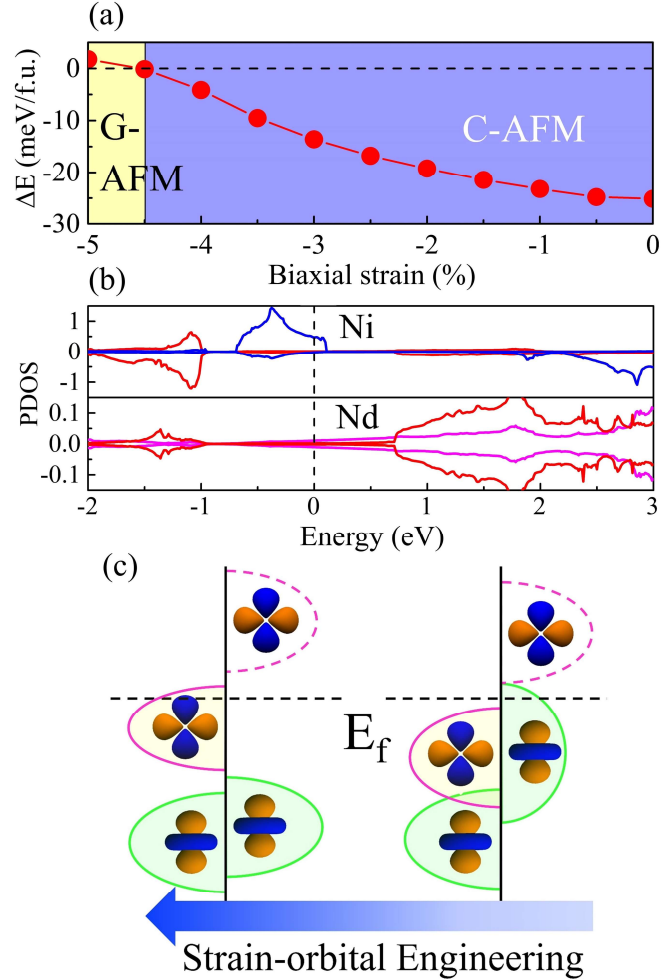

FIG. 8. Triggering cuprate-like electronic and magnetic properties by strain-orbital engineering. (a) Epitaxial-strain-dependent magnetic phase diagram of NdNiO2 thin film and (b) the PDOS of Nd and Ni for NdNiO2 under 5% compressive strain. (c) Schematic representation of the mechanism for strain triggered electronic and magnetic phase transitions.

Strain engineering has long been demonstrated to be an effective approach to modulate the on-site energy of orbitals and eventually tailor the related properties [69,70]. In ABO3 perovskites, compressive epitaxial biaxial strain directly leads to the elongation (contraction) of the out-of-plane (in-plane) lattice constant, which typically weakens (strengthens) the hybridization of $d_{3z^2-r^2}$ ($d_{x^2-y^2}$) orbital with the surrounding O $p$ orbitals, thus shifting the on-site energy of $d_{3z^2-r^2}$ ($d_{x^2-y^2}$) orbital to a lower (higher) level. To verify whether this scenario could be applied to infinite-layer NdNiO$_2$, biaxial strain dependence of energy difference ΔE between C-AFM and G-AFM phases is firstly explored and represented in Fig. 8(a). We see that ΔE gradually decreases upon increasing the compressive strain and there is a crossover from C-AFM to G-AFM order at around 4.5% compressive strain. To verify the above qualitative analysis, we calculate the exchange constant to provide a quantitative picture. We found that the out-of-plane exchange constant is reduced to about -0.87 meV and the nearest-neighbor in-plane exchange constant is about 37 times larger than the nearest-neighbor out-of-plane exchange constant, highlighting a strong quasi-2D magnetic coupling.

As we mentioned above, the appearance of out-of-plane exchange interaction is connected to the spin-polarized itinerant $d_{3z^2-r^2}$ electron. It is obvious from the PDOS, displayed in Fig. 8(b), that the spin polarization of $d_{3z^2-r^2}$ orbital almost vanishes when compressive strain reaches 5% and thereby yields the quasi-2D AFM order. Moreover, we see that the band edge of $d_{x^2-y^2}$ orbital goes across the Fermi level and governs the states around the Fermi level. The design principle of strain engineering is summarized in Fig. 8(c), a crossover from original 3D to cuprate-like 2D magnetic dimensionality dominated by $d_{x^2-y^2}$ orbital is triggered. The strained NdNiO2 thin film then obviously combine the key ingredients of CaCuO2 (see Figs. 4(a) and 5(a)): strong quasi-2D AFM order, almost pure single $d_{x^2-y^2}$ band near the Fermi level and large orbital polarization between the $d_{x^2-y^2}$ and $d_{3z^2-r^2}$ orbitals.

## VII. OUTLOOK

Our first- and second-principles results highlight that, despite sharing the same reference atomic structure, $R$NiO$_2$ infinite-layer compounds differ significantly from high-T$_c$ cuprate superconductors such as CaCuO$_2$. Our calculations unveil that, contrarily to CaCuO$_2$, $R$NiO$_2$ (with $R$=Pr-Lu) compounds are prone to develop AFD oxygen-rotation distortions similarly to $R$NiO$_3$ perovskites and exhibit a displacive phase transition from *P4/mmm* to *I4/mcm* at low temperature.

These AFD oxygen-rotation motion appear to be the key to provide a consistent description of the structural, electronic and magnetic properties of infinite layers. At first, AFD oxygen rotation motion simultaneously localize the electron and suppress the self-doping effect from Nd atom, which can explain the unusual upturn of resistivity observed at low temperature [1]. The rare-earth ion rotation evolution, the rotation controlled bandwidth and the magnetic coupling strength are completely in line with $R$NiO$_3$ perovskites. Additionally, unlike CaCuO$_2$, we emphasize that the Ni $d_{3z^2-r^2}$ orbital dominates the states at the Fermi level. The strong k$_z$ dispersion of itinerant $d_{3z^2-r^2}$ electron is not only critical in determining the strong interlayer magnetic coupling and 3D features of exchange interactions (that contrast with 2D features in CaCuO$_2$), but is also relevant to explain that the upper critical field is surprisingly isotropic at low temperature [4,5]. At the electronic and magnetic levels, $R$NiO$_2$ compounds are therefore closely similar to ferropnictides [50,51,58-60]. As a summary of our calculations, we then successfully built a complete phase diagram, summarizing the temperature evolution of structural and magnetic phases in the whole family of $R$NiO$_2$ compounds (Fig. 7).

Having revealed the similarities between $R$NiO$_3$ perovskites and $R$NiO$_2$ infinite layers, it naturally appears that the strategies currently used to control the properties of perovskites (rotation engineering [35,71], strain engineering [72], $R$-site element substitution [73], and interfacial coupling [74]) could be exploited in infinite layers to achieve the continuous control of their superconducting properties. Relying then on strain-orbital engineering, we highlight that applying epitaxial compressive strains, as practically achievable in thin films, it is possible to shift the energy of Ni $d_{x^2-y^2}$ ($d_{3z^2-r^2}$) orbital to a higher (lower) level and to trigger a magnetic transition from the 3D C-AFM phase to a 2D G-AFM phase. Such a strained NdNiO$_2$ now closely resembles CaCuO$_2$ -- showing strong orbital polarization of two *e$_g$* orbitals with the $d_{x^2-y^2}$ orbital dominating the Fermi surface and strong quasi-2D AFM correlations -- and might show improved superconducting properties.

In conclusion, our work not only provides a deeper fundamental understanding of $R$NiO$_2$ compounds that explicitly combines the interplay between their structural, electronic and magnetic degrees of freedom but also exploits this knowledge to provide new pathways to improve their properties. As such, we hope that our work will motivate experimentalists to further exploit the coupling between rotation, charge, orbital, spin, and strain in infinite layer $R$NiO$_2$ and other ABX$_2$ system to realize the practical control of their properties.

## VIII. METHODS

We perform first-principles calculations using density functional theory (DFT) for the ground state properties calculations, finite-temperature second-principles model to calculate the structural phase transition, TB2J code [61] to extract orbital-dependent exchange constant from maximally localized Wannier functions (MLWF) and vampire code to calculate the Néel temperature [75].

**A. First-principles calculations.**

Our first-principles calculations are performed using the projector augmented wave (PAW) method implemented in VASP code [76,77]. Recent work shows that the electronic and magnetic properties of superconducting La$_2$CuO$_4$ is significantly improved [49] with SCAN [34] functional. Thus, widely used GGA functional [32] is employed for lattice dynamical analysis and we employ the SCAN for calculating of electronic and magnetic properties. Results from GGA and GGA + U [33] are also compared and discussed (see Appendix C). For SCAN functional, an energy cutoff of 900 eV is used, the force and energy convergence criterions are set to be $10^{-8}$ eV and $10^{-4}$ eV/Å for the structural relaxation. Lower kinetic energy of 700 eV and force and energy convergence criterions of $10^{-7}$ eV and $10^{-3}$ eV/Å are used for GGA and GGA + U. The kpoint meshes [78] are set to be $9 \times 9 \times 7$ and $6 \times 6 \times 7$ for $\sqrt{2} \times \sqrt{2} \times 2$ and $2 \times 2 \times 2$ supercells.

## B. Second-principles calculations

First-principles based second-principles method is applied to investigate the evolution of rotation motion at different temperature. We build a second-principles model with displacements of oxygen atom $\mu$ as the degree of freedom. The total energy can be expressed as:

$$E_{total} = E_{self} + E_{short} \qquad (1)$$

Where $E(\mu_i)$ in the first term $E_{self} = \sum_i E(\mu_i)$ represents the energy of an isolated oxygen atom at $i$-th location with amplitude $\mu_i$. It's truncated at fourth order since we found that's enough to describe the PESs. Due to the symmetry consideration, $E(\mu_i)$ can be written as:

$$E(\mu_i) = k_1 \mu_i^2 + k_2 \mu_i^4 \qquad (2)$$

Where $k_1$ and $k_2$ refers to the parameters to be determined from first-principles calculations. In the second term $E_{short} = \sum_{i,j} j_{ij} \mu_i \mu_j$, $\mu_i$ and $\mu_j$ are the amplitude of oxygen atom displacements at $i$-th and $j$-th location and $j_{ij}$ is the coupling parameter between them. This term is the energy contribution from the short-range interactions between neighboring oxygen atoms. The length of short-range interactions is truncated at one unit cell, due to the symmetry consideration, there are only four independent interaction parameters ($j_1$, $j_2$, $j_3$ and $j_4$) for the short-range interactions as sketched in Appendix D, Fig. 19(a). All these parameters are obtained from first-principles calculations.

Based on the second-principles model, the Monte Carlo simulations are performed to investigate the rotation motion at different temperature. The total energies obtained from first-principles calculations and second-principles model are compared in Appendix D, Fig. 19(b). The close match of energy guarantees the accuracy and validity of our model.

## C. Exchange constant calculations

The exchange constants and the orbital contributions are obtained by the magnetic force theorem [79] as implemented in the TB2J code [61] using maximally localized Wannier functions [80,81]. Here, Nd $d_{xy}$ and $d_{3z^2-r^2}$ orbitals, Ni $d_{xy}$, $d_{yz}$, $d_{zx}$, $d_{x^2-y^2}$, and $d_{3z^2-r^2}$ orbitals and O $p_x$, $p_y$, and $p_z$ orbitals are used for the construction of Wannier functions.

**ACKNOWLEDGMENTS**

This work is supported by the Initial Scientific Research Fund of Lanzhou University for Young Researcher Fellow (Grant No. 561120206) and the National Natural Science Foundation of China (Grant No. 051B22001). The calculations are supported by the Center for Computational Science and Engineering of LanZhou University and Southern University of Science and Technology.

## APPENDIX A: COMPENSATION OF THE STRUCTURAL PROPERTIES OF $R$NiO₃, $R$NiO₂ and CaCuO₂

### 1. Unstable modes in LuNiO₂

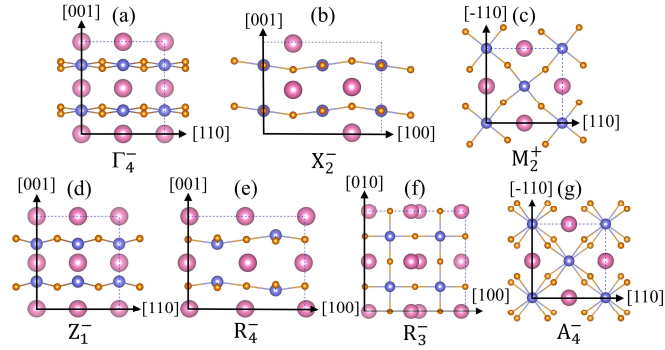

FIG. 9. Schematic pictures of the unstable modes in LuNiO₂ including (a) $\Gamma_4^-$, (b) $Z_1^-$, (c) $R_4^-$, (d) $R_3^-$, (e) $X_2^-$, (f) $M_2^+$ and (g) out-of-phase rotation $A_4^-$.

### 2. Phonon dispersion comparison in $R$NiO₃

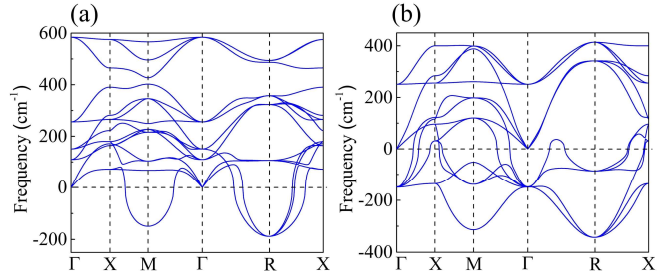

FIG. 10. Phonon dispersions of the ferromagnetic Pm$\bar{3}$m (a) LaNiO₃ and (b) LuNiO₃.

Fig. 10 plots the phonon dispersions of the ferromagnetic Pm$\bar{3}$m LaNiO$_3$ and LuNiO$_3$. The unstable mode with the largest imaginary frequency at high-symmetry R point corresponds to the out-of-phase rotation mode. The imaginary frequencies related to the M-point in-phase rotation and R-point out-of-phase rotation are obviously increased in LuNiO$_3$. Moreover, more unstable modes appear in LuNiO$_3$.

### 3. Strain induced rotation in LaNiO$_2$ and CaCuO$_2$

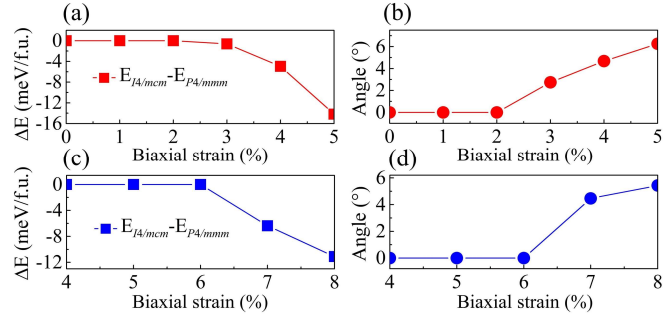

FIG. 11. (a) Energy difference between G-AFM *I4/mcm* and G-AFM *P4/mmm* phase and (b) rotation angle as a function of strain in the ground state structure for LaNiO$_2$. The energy difference between G-AFM *I4/mcm* and G-AFM *P4/mmm* phase and strain dependent rotation angle of CaCuO$_2$ thin film are shown in (c) and (d), respectively. The results are obtained from GGA functional.

Fig. 11 shows the *P4/mmm* - *I4/mcm* transition for LaNiO$_2$ (Figs. 11(a) and 11(b)) and CaCuO$_2$ (Figs. 11(c) and 11(d)) thin films, respectively. Sizable out-of-phase rotation could be induced in both thin films by epitaxial tensile strain. The results suggest that rotation maybe a common distortion in ABX$_2$ systems similar with ABX$_3$ perovskites.

### APPENDIX B: FERMI SURFACE OF NdNiO$_2$

Fig. 12 compares the Fermi surface of *P4/mmm* and *I4/mcm* NdNiO$_2$, it is obvious that the elliptical Fermi surface of the ground state *I4/mcm* phase at the Γ point induced by the self-doping effect of Nd is strongly reduced compared with the high-symmetry *P4/mmm* phase.

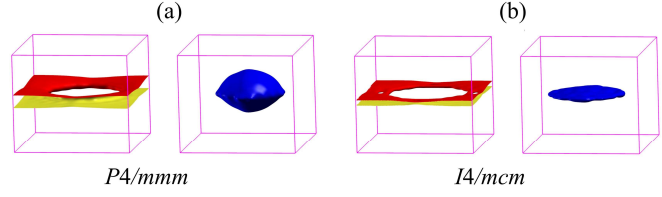

FIG. 12. Fermi surface of (a) *P4/mmm* and (b) *I4/mcm* NdNiO$_2$.

**APPENDIX C: REPRODUCE OF THE MAIN RESULTS OF SCAN FUNCTIONAL BY GGA + U METHOD**

Here, we emphasize that, while our results from SCAN functional provide new and remarkably different electronic and magnetic insight comparing with previous first-principles calculations, it is worth mentioning that the results are not an accidental finding of specific SCAN functional, all the main ground state features can be well reproduced by the DFT +U method with a careful examination of U.

**1. C-AFM magnetic ground state**

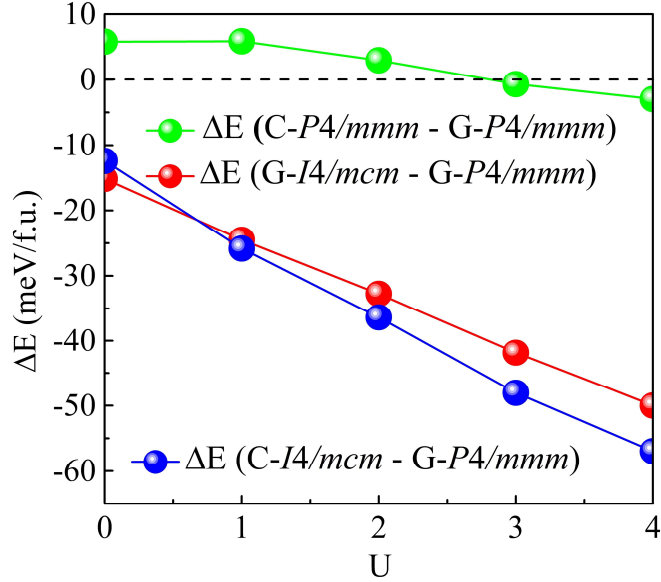

FIG. 13. Energy difference of C-AFM *P4/mmm* phase, C-AFM *I4/mcm* phase and G-AFM *I4/mcm* phases relative to the reference G-AFM *P4/mmm* phase for bulk NdNiO$_2$.

Fig. 13 compares the energy of *P4/mmm* and *I4/mcm* NdNiO$_2$ in G-AFM and C-AFM orders, *I4/mcm* phase always has lower energy than the *P4/mmm* phase for GGA and GGA + U methods, the larger the U values, the larger the energy difference between two phases. Note that G-AFM is the magnetic ground state for GGA functional, while C-AFM becomes the ground state when Hubbard U equals to 1-3 eV.

### 2. Orbital-selective Mott physics and weakened metallicity

Fig. 14 shows the PDOS of *P4/mmm* and *I4/mcm* NdNiO$_2$ from GGA + U (U = 1-3 eV) method. It is obvious that the dominate states at Fermi level using GGA + U are Ni $d_{3z^2-r^2}$ electron which resembles the characters of SCAN results. Orbital-selective Mott localization with the $d_{3z^2-r^2}$ itinerant electron band and $d_{x^2-y^2}$ local electron band is also similar with the results of SCAN. Additionally, the reduction of self-doping from Nd $d_{3z^2-r^2}$ orbital and bandwidth of Ni $d_{x^2-y^2}$ orbital due to the appearance of rotation are also clearly reproduced by GGA + U results.

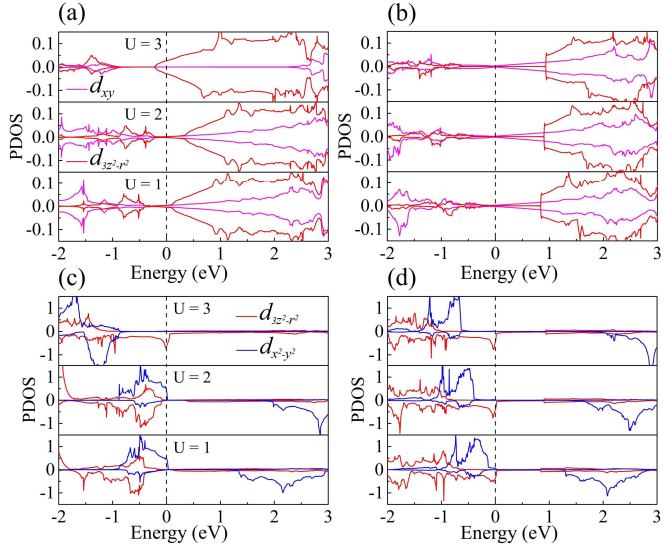

FIG. 14. PDOS of Nd atom in the (a) high-symmetry *P4/mmm* and (b) low-symmetry *I4/mcm* NdNiO$_2$ phases, the corresponding PDOS of Ni atom in the high-symmetry and low-symmetry phases are shown in (c) and (d), respectively. In each figure, the U value increases from 1 to 3 eV from the bottom to the top.

### 3. 3D magnetic interactions

Fig. 15 shows the in-plane and out-of-plane nearest-neighbor exchange constant as the U value changes from 1 to 3 eV. we can see a clear 3D feature of exchange interactions in NdNiO2 with notable interlayer ferromagnetic coupling.

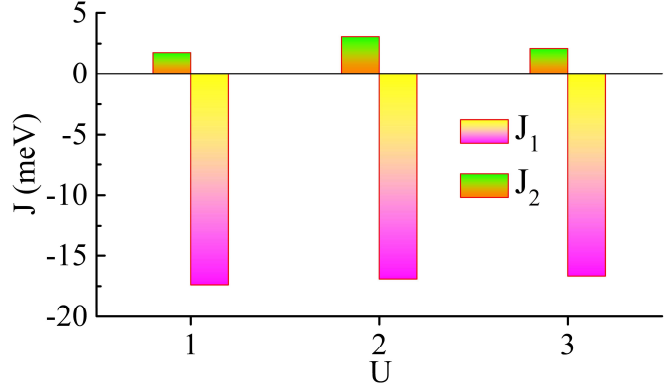

FIG.15. In-plane and out-of-plane nearest-neighbor exchange constant of NdNiO$_2$ with different U values (U = 1-3 eV).

## 4. Strain triggered 3D to 2D electronic and magnetic phase transitions

Fig. 16 shows strain dependent magnetic phase diagram calculated by GGA + U (U = 1-3 eV) methods. Strain triggered C-AFM phase to G-AFM phase transition is clearly reflected.

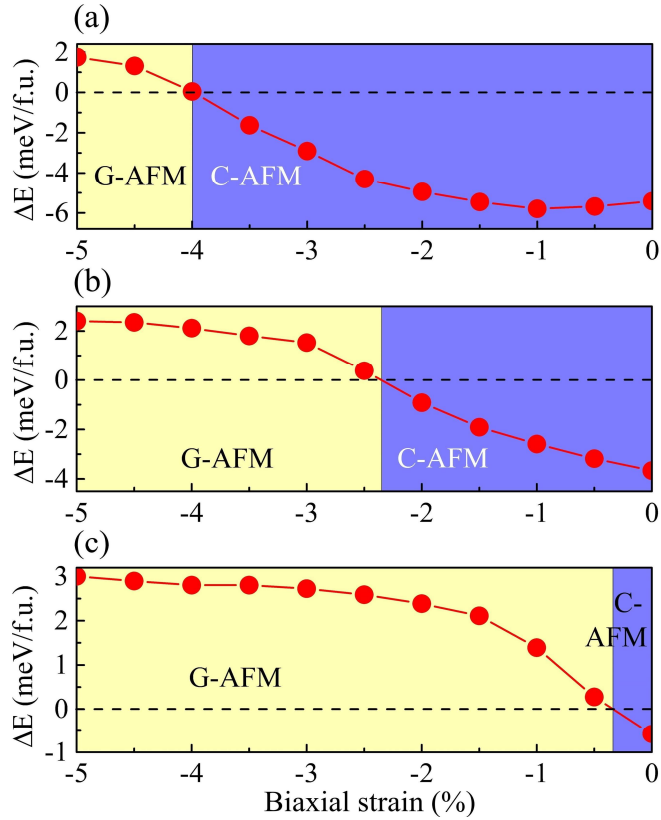

FIG. 16. Strain-magnetic phase diagram of NdNiO$_2$ thin film using the GGA + U with (a) U = 3 eV, (b) U = 2 eV and (c) U = 1 eV.

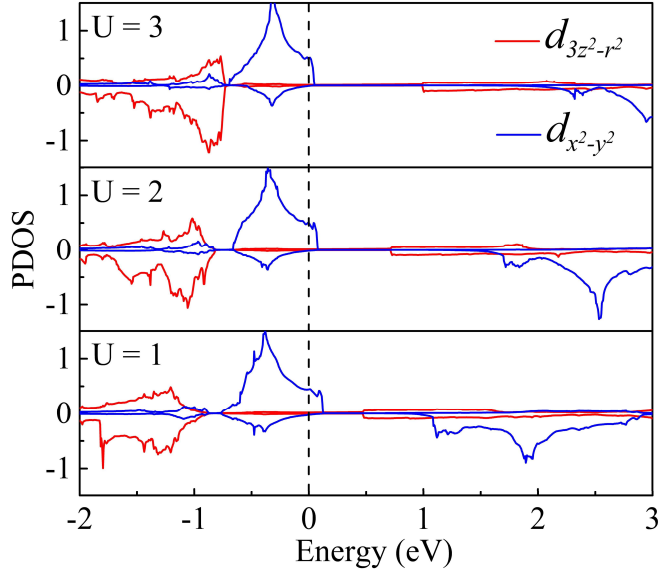

FIG. 17. PDOS of Ni atom of NdNiO$_2$ thin film under 5% compressive strain with GGA + U method.

Fig. 17 shows the PDOS of NdNiO$_2$ thin film under 5% compressive strain from GGA + U. Obviously, the dominant states at the Fermi level have changed from $d_{3z^2-r^2}$ band (see Fig. 14(d)) to $d_{x^2-y^2}$ band in line with the SCAN results.

Fig. 18 compares the ratio of in-plane and out-of-plane nearest-neighbor exchange constant of bulk CaCuO$_2$ and NdNiO$_2$, and NdNiO$_2$ thin film under 5% compressive strain. From both SCAN and GGA + U method, it is obvious the magnetic anisotropy is greatly enhanced in the thin film. The ratio of in-plane and out-of-plane nearest-neighbor exchange constant in the strained thin film is comparable with 2D CaCuO$_2$, indicating the strong quasi-2D feature of magnetic coupling.

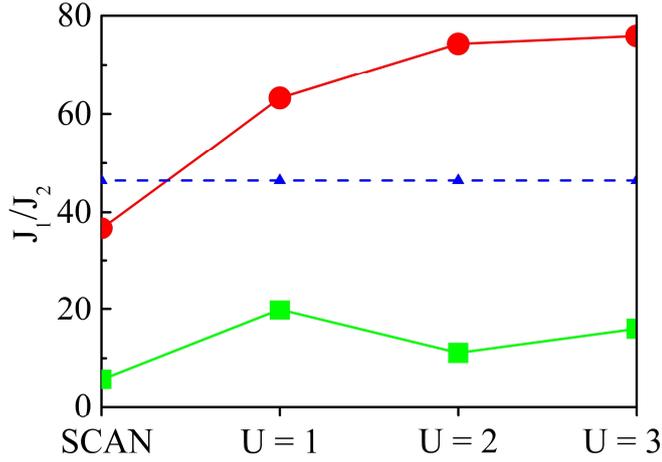

FIG. 18. The ratio of in-plane and out-of-plane nearest-neighbor exchange constant of bulk $NdNiO_2$ (green curve) and thin film under 5% compressive strain (red curve) from SCAN and GGA + U (U = 1-3 eV). The ratio of $CaCuO_2$ is also shown in the dashed line for comparison.

Consequently, strain controlled 3D to 2D magnetic dimensionality transition is a general feature of $NdNiO_2$ thin film when SCAN functional and GGA + U method are employed.

**APPENDIX D: ACCURACY OF THE SECOND-PRINCIPLES MODEL**

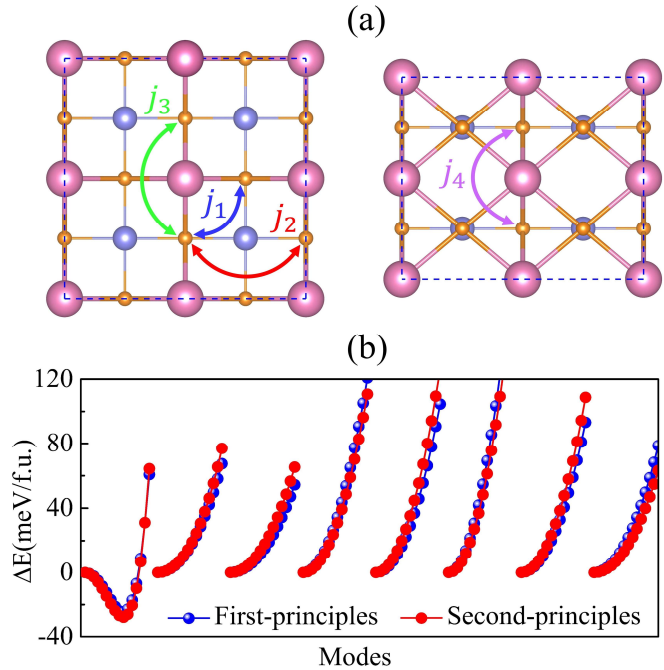

FIG. 19. (a) Schematic representation of four short-range interactions used in the second-principles model, (b) the comparison of the energy evolutions between first-principles calculations and second-principles model for eight different modes used for fitting parameters in the second-principles model. The first mode is the out-of-phase rotation.